\UseRawInputEncoding
\documentclass[%
 reprint,
superscriptaddress,
 amsmath,amssymb,
 aps,
prb,
]{revtex4-1}

\usepackage{graphicx}
\usepackage{dcolumn}
\usepackage{bm}
\usepackage{amsmath,amssymb}
\usepackage{graphicx}
\usepackage{color}

\usepackage{hyperref}

\begin{document}

\preprint{AIP/123-QED}

\title{Hybrid Au-Si microspheres produced by laser ablation in liquid (LAL) for temperature-feedback optical nano-sensing and anti-counterfeit labeling$^\dag$}

\author{S.~Gurbatov}
\affiliation{Far Eastern Federal University, Vladivostok 690090, Russia}
\affiliation{Institute of Automation and Control Processes, Far Eastern Branch, Russian Academy of Science, Vladivostok 690041, Russia}

\author{V.~Puzikov}
\affiliation{Far Eastern Federal University, Vladivostok 690090, Russia}
\affiliation{Institute of Automation and Control Processes, Far Eastern Branch, Russian Academy of Science, Vladivostok 690041, Russia}

\author{D.~Storozhenko}
\affiliation{Institute of Automation and Control Processes, Far Eastern Branch, Russian Academy of Science, Vladivostok 690041, Russia}

\author{E.~Modin}
\affiliation{CIC nanoGUNE BRTA, E-20018 Donostia - San Sebastian, Spain}

\author{E.~Mitsai}
\affiliation{Institute of Automation and Control Processes, Far Eastern Branch, Russian Academy of Science, Vladivostok 690041, Russia}

\author{A.~Cherepakhin}
\affiliation{Institute of Automation and Control Processes, Far Eastern Branch, Russian Academy of Science, Vladivostok 690041, Russia}

\author{A.~Shevlyagin}
\affiliation{Institute of Automation and Control Processes, Far Eastern Branch, Russian Academy of Science, Vladivostok 690041, Russia}

\author{A.V.~Gerasimenko}
\affiliation{Institute of Chemistry FEB RAS, Vladivostok 690022, Russia}

\author{S.A.~Kulinich}
\affiliation{Far Eastern Federal University, Vladivostok 690090, Russia}
\affiliation{Research Institute of Science and Technology, Tokai University, Hiratsuka, Kanagawa 259-1292, Japan}

\author{A.~Kuchmizhak}
\email{alex.iacp.dvo@mail.ru}
\affiliation{Institute of Automation and Control Processes, Far Eastern Branch, Russian Academy of Science, Vladivostok 690041, Russia}
\affiliation{Pacific Quantum Center, Far Eastern Federal University, 8 Sukhanova Str., Vladivostok, Russia}

\begin{abstract}
Recent progress in hybrid nanomaterials composed of dissimilar constituents permitted to improve performance and functionality of novel devices developed for optoelectronics, catalysis, medical diagnostic and sensing. However, the rational combination of such contrasting materials as noble metals and semiconductors within individual hybrid nanostructures $via$ a ready-to-use and lithography-free fabrication approach is still a standing challenge. Here, we report on a two-step synthesis of Au-Si microspheres generated by laser ablation of silicon in isopropanol followed by laser irradiation of the produced Si nanoparticles in presence of HAuCl$_4$. Thermal reduction of [AuCl$_4$]$^-$ species to metallic gold phase, along with its subsequent mixing with silicon under laser irradiation creates a nanostructured material with a unique composition and morphology revealed by electron microscopy, tomography and elemental analysis. Combination of such basic plasmonic and nanophotonic materials as gold and silicon within a single microsphere allows for efficient light-to-heat conversion, as well as single-particle SERS sensing with temperature feedback modality and expanded functionality. Moreover, the characteristic Raman signal and hot-electron induced nonlinear photoluminescence coexisting within Au-Si hybrids, as well as commonly criticized randomness of the nanomaterials prepared by laser ablation in liquid, were proved useful for realization of anti-counterfeiting labels based on physically unclonable function approach.

\end{abstract}

\maketitle

\section{Introduction}

Resonant interaction of optical radiation with the matter is in the heart of modern sensors, as well as of light-emitting and optoelectronic devices. Coupling of light waves with resonant oscillations of free electron plasma supported by noble-metal nanostructures and metal-dielectric interfaces (i.e. localized and propagating surface plasmons \cite{maier2007plasmonics}) provides an efficient way to boost this interaction, giving rise to a number of practically important interface phenomena and their related applications, such as efficient light-to-heat conversion, surface-enhanced Raman scattering (SERS), plasmon-mediated generation of hot electrons, etc \cite{sobhani2013narrowband,baffou2020applications,langer2019present}. An alternative way recently suggested to advance light-matter interactions consists in the use of nanophotonic (all-dielectric) platforms made of optically resonant dielectric or semiconductor materials that exhibit high refractive index and low optical losses (e.g., Si, Ge, TiO$_2$, etc.) \cite{kuznetsov2016optically}. Related nanostructures and their ordered arrangements, i.e. metasurfaces, permit to achieve full control over the properties of scattered/ reflected/ transmitted light waves, offering resonant light absorption and enhanced light emission/ nonlinear effects \cite{yu2014flat,koshelev2020subwavelength,koshelev2020dielectric}. These features are partially related to the ability of all-dielectric nanostructures to confine incident radiation inside their bulk, which cannot be achieved with their lossy plasmonic counterparts.

Due to a limited choice of basic materials with on-demand properties, integration of plasmonic and nanophotonic approaches within hybrid nanostructures is considered a key strategy for designing next-generation optoelectronic devices and optical sensors \cite{soukoulis2011past,jiang2014metal,makarov2017light,fusco2020photonic}. Meanwhile, along with their ability to integrate such contrasting materials as noble metals and semiconductors at the nanoscale, the related fabrication approaches should meet simultaneously a number of practical requirements such as scalability, simplicity, attractive production yield, etc. Beyond the multi-step and expensive lithography-based approaches, laser-assisted technologies have recently appeared as a promising route towards environment friendly high-yield preparation of hybrid nanomaterials \cite{zeng2012nanomaterials,zhang2017laser,amendola2020room,shabalina2022green}. In particular, laser ablation in liquid (LAL) presents a flexible experimental approach that utilizes intense laser pulses to ablate a target material placed in a liquid medium. Spatially and temporally confined laser radiation can be used to create unique experimental conditions (high pressures and temperature, fast quenching rates), also driving  chemical interactions of ejected nanomaterial with surrounding liquid. This provides a pathway for the preparation of nanomaterials with a wide range of morphology, structure and composition (including meta-stable and non-equilibrium phases). Laser irradiation of LAL-generated dispersions can further expand the method functionality, allowing for high-performance preparation of diverse hybrid nanomaterials for photovoltaics \cite{tarasenka2021laser,furasova2021mie}, photothermal conversion \cite{nastulyavichus2020multifunctional,gurbatov2021black}, catalysis \cite{zhang2020laser,forsythe2021pulsed,shabalina2022laser}, nonlinear optics \cite{gurbatov2022hybrid}, sensing \cite{amendola2014magneto,scaramuzza2017magnetically,bharati2019explosives,guadagnini2021kinetically} and medical applications \cite{torresan20204d,amendola2021polymer}.

Gold, the most chemically stable plasmon-active metal, and silicon, an earth-abundant semiconductor widely applied for all-dielectric metasurface design, present an intriguing combination, in which the advantages of plasmonic and nanophotonic concepts can be merged within unified nanostructures. Although extensive research was conducted and reported over years on products of Au and Si ablation in various liquids  \cite{simakin2004nanoparticles,amendola2009laser,al2018recent,zabotnov2020nanoparticles}, few papers reported on formation of Au-Si hybrids with various morphologies (nano-alloys, core-satellites, etc.\cite{liu2015fabrication,saraeva2018laser,ryabchikov2019facile,kutrovskaya2017synthesis,al2021laser}). However, no general assessment of their nanophotonic properties and/or demonstration of material's suitability for related applications was shown. Here, hybrid Au-Si microspheres (MSs) were produced $via$ ablation of Si wafer by nanosecond (ns)-pulsed laser in isopropanol followed by further irradiation of the as-produced Si dispersion in presence of HAuCl$_4$. Formation of Au nanoparticles on the surface of Si particles through the thermal reduction of [AuCl$_4$]$^-$ species, as well as further mixing of the metal and semiconductor phases upon remelting and recrystallization, were found to result in a novel hybrid Au-Si product. By combining transmission electron microscopy (TEM), electron tomography, energy-dispersive X-ray spectroscopy (EDX) elemental analysis, X-ray diffraction (XRD) and Raman spectroscopy, we revealed the unique structure of the newly prepared hybrids, in which nanocrystalline Si inclusions are wrapped and decorated with nanostructured Au. Strong light absorbing properties, characteristic Raman signal and nonlinear hot-electron-induced photoluminescence coexisting in the novel Au-Si MSs were proved to be useful for realization of single-particle SERS sensors with temperature-feedback modality and fabrication of unclonable anti-counterfeit labels.

\begin{figure*}[t!]
\centering
\includegraphics[width=1.8\columnwidth]{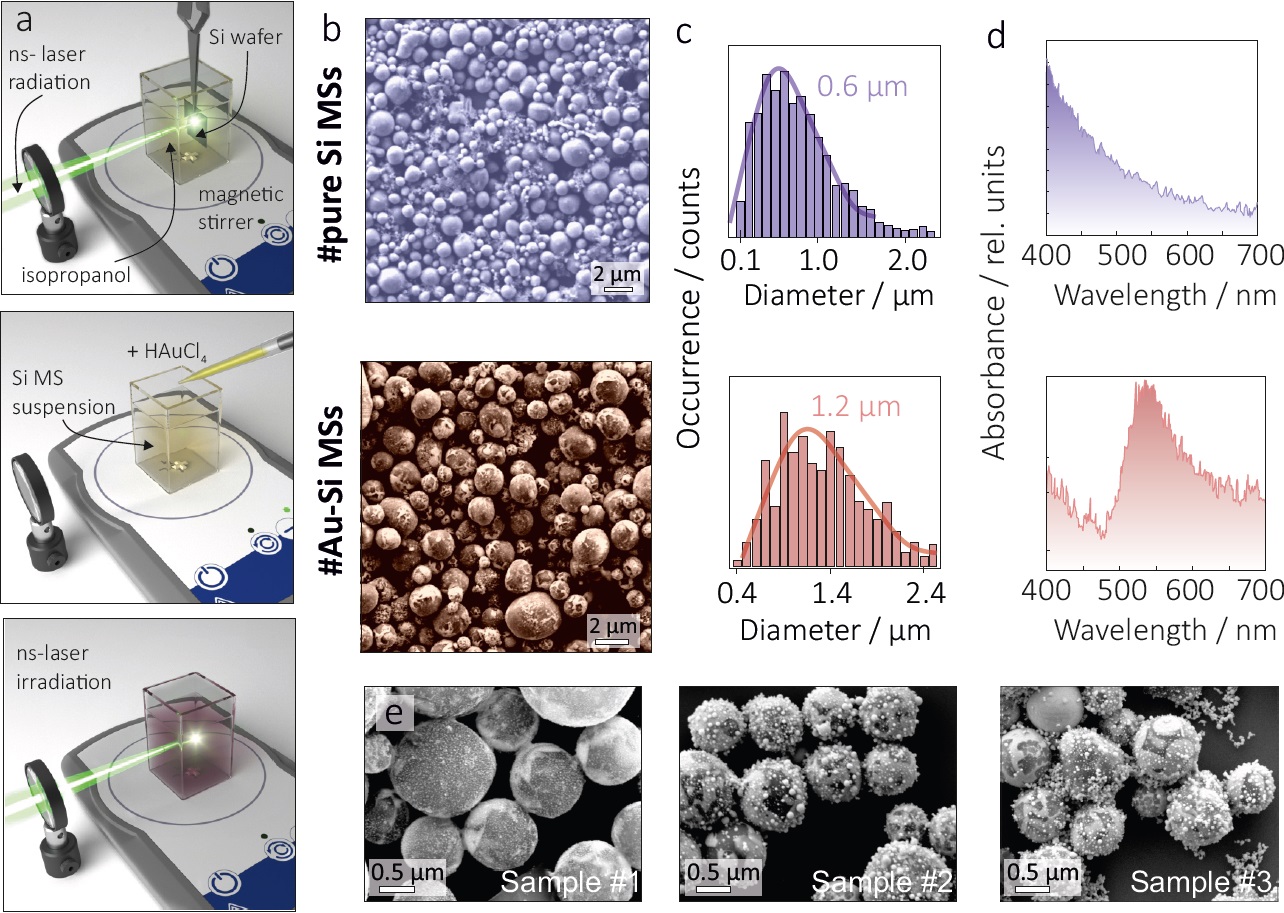}
\caption{\textbf{Fabrication of Au-Si MSs.} (a) Schematic presentation of preparation of Au-Si hybrid structures $via$ LAL. (b) Representative top-view SEM images, (c) size distributions and (d) normalized UV–vis absorbance spectra (isopropanol suspensions) of Si precursor (top row) and hybrid Au-Si MSs (bottom row, Sample \#2). (e) SEM images of Au-Si MSs produced by varying the amount of HAuCl$_4$ (0.25, 0.5 and 0.75 mM) added to laser-treated suspension.}
\label{fig:1}
\end{figure*}

\section{Results and discussion}
\subsection{Preparation of Au-Si microspheres $via$ LAL}
Hybrid Au-Si microstructures were produced using a simple two-step protocol schematically illustrated in Figure 1a (see also Methods section). Briefly, a dispersion with Si MSs with an average diameter of $\approx$0.6 $\mu$m was first produced by ablating a bulk monocrystalline Si wafer immersed in isopropanol (Figure 1b). Next, a certain amount of aqueous HAuCl$_4$ (10$^{-3}$M) was added to the as-produced dispersion with Si MSs followed by its laser-irradiation for an additional 1 h using the same LAL parameters (pulse energy of 0.4 mJ and pulse repetition rate of 20 Hz). The resultant Au-Si product obtained after adding 0.5 mL of HAuCl$_4$ solution is shown in the SEM image (Figure 1b) revealing the average particle diameter of $\approx$1.2 $\mu$m. Comparative UV-vis absorbance spectra of the dispersions containing the initial Si precursor and the Au-Si product are presented in Figure 1c. The spectrum of the product is seen to reveal a characteristic absorption band at 530 nm, which confirms appearance of metallic Au phase formed $via$ thermal reduction processes.

To demonstrate the ability to control the composition of the produced material, we first fixed the laser irradiation conditions varying only the amount of HAuCl$_4$ (0.25, 0.5 and 0.75 mL) added to the processed liquid. UV-vis spectra of the three produced dispersions (hereafter referred to as Samples \#1, \#2 and \#3) all demonstrated absorption band at 530 nm whose intensity gradually increased along with the amount of HAuCl$_4$ added during preparation. This clearly indicates the increase in the content of Au within the product structures (Figure S1, ESI). The resultant Au-Si materials (Samples \#1 to \#3) were then taken from each dispersion and drop-cast on Si wafer for further analysis. Close-up SEM images of Au-Si MSs produced at three different concentrations of HAuCl$_4$ are shown in Figure 1e, revealing complex morphology and composition of the obtained product. It is worth noting that Si and Au elements demonstrate remarkably different contrast when visualized by SEM with any secondary or back-scattered electrons detector. This  permits to distinguish both elements within the bulk of spherically shaped MSs (mainly in Samples \#2 and \#3). Moreover, nano-sized Au particles decorating Si-based MS surface are also clearly identified in the SEM images of all samples.

\begin{figure*}
\centering
\includegraphics[width=1.9\columnwidth]{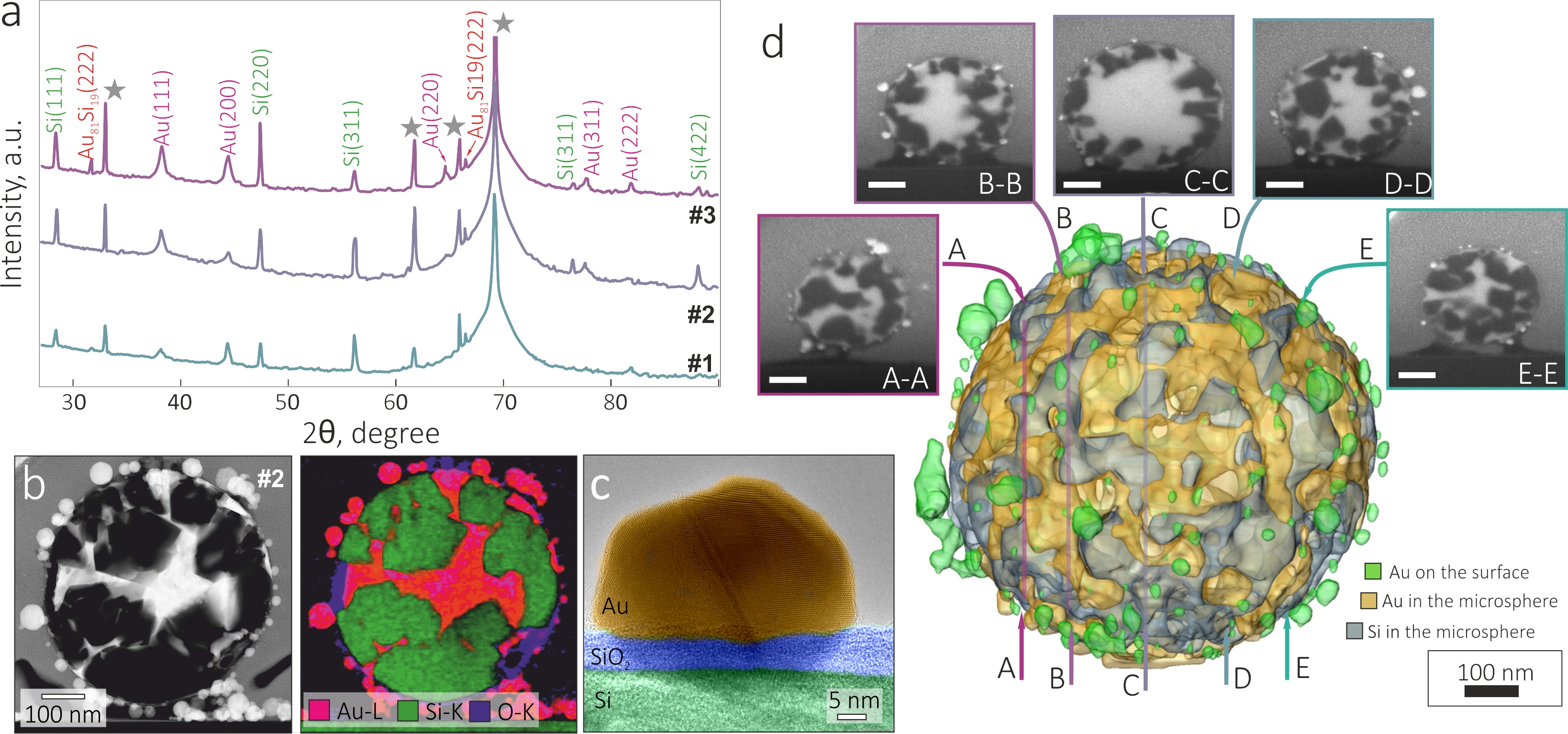}
\caption{\textbf{Characterization of Au-Si MSs}. (a) XRD patterns of Au-Si MSs produced at different concentrations of HAuCl$_4$ added to laser-processed dispersion (Samples \#1 - \#3). Stars denote XRD peaks attributed to the Si substrate used. (b) TEM image and EDX composition mapping of a representative Au-Si MS (Sample \#2). (c) HR-TEM image of an isolated Au nanoparticle on the surface of Au-Si MS, with a well-seen SiO$_2$ shell. (d) 3D model of an isolated Au-Si MS reconstructed from a series of 50 cross-sectional FIB cuts. Several representative cuts are shown as insets. Au nanoparticles on the surface of a Au-Si MS are highlighted by green color to distinguish them from Au material (orange) mixed with Si within the particle.
}
\label{fig:1}
\end{figure*}

Analysis of SEM images shows that the increase of the amount of HAuCl$_4$ added to laser-treated mixture results in a visually larger amount of Au (both inside the MSs and on their surface). This finding was confirmed by systematic EDX studies of as-prepared hybrid MSs (Figure S2, ESI), showing the average content of Au (in wt.\%) gradually increases from 26$\pm$1 (Sample \#1) to 48$\pm$2.5 wt.\% (Sample \#3). EDX characterization also revealed almost constant content of carbon atoms ($\approx$ 4.5$\pm$0.5 wt.\%) in the product, which presumably resulted from decomposition of isopropanol molecules upon laser irradiation. In this study, alcohol medium was used to minimize Si oxidation during material preparation. According to EDX analysis, the oxygen content increased from 3$\pm$0.5 (Sample \#1) to 5.2$\pm$0.7 wt.\% (Sample \#3) which could also originate from  laser-induced Si oxidation stimulated by adding aqueous HAuCl$_4$.

The same Au-Si MSs deposited on Si(001) substrate (Samples \#1-\#3) were further systematically analyzed by XRD (see Methods section) revealing several key features regarding the composition and crystal structure of the LAL-prepared hybrids. First of all, clear signatures of both cubic Au (Fm3m, JCPD04-0784) and Si (Fd3m, JCPD27-1402) can be identified in the corresponding XRD patterns of all samples with no signs of metastable Si phases often associated with LAL processing \cite{liu2013violet}. The intensity of the main Au-related peaks (2$\theta\approx$ 38.1$^o$, 44.4$^o$, and 81.8$^o$) was found gradually to increase with the amount of HAuCl$_4$ added to the processed liquid, thus confirming the EDX measurement results. Moreover, at higher HAuCl$_4$ addition, a larger number of peaks related to Au (2$\theta\approx$ 64.7$^o$ and 77.6$^o$) are seen in Figure 2a, as well as peaks of Au silicide with composition Au$_{81}$Si$_{19}$ (seen at 2$\theta\approx$ 32.2$^o$; 39.7$^o$ and 67.1$^o$; JCPD39-0735, Sample \#3).

  Also, at higher HAuCl$_4$ concentration in the LAL-processed liquid, the most-intense Au-related XRD peak changes from Au(200) to Au(111), indicating that the Au-Si system tends to the most favorable orientation of Au(111)||Si(111) which is resulted from the minimum crystal lattice mismatch between Au particles and their Si support \cite{langford1978scherrer}. The average grain size of Si phase in the MSs $D_{g}$ calculated using Scherrer's equation  $D_{g}$=(K$\cdot\lambda_{X-ray}$)/($\beta\cdot$cos$\theta$) (K is a constant depending on grain shape and crystal symmetry, $\lambda_{X-ray}$ is the X-ray wavelength, $\beta$ is the FWHM of the diffraction peak, and $\theta$ is the diffraction angle \cite{daudin2012epitaxial}) was found to be $\approx$ 125$\pm$10 nm and nearly same for all tested samples. The lattice of Si phase was found to be completely relaxed as its interplanar distances corresponded to those of bulk Si, with residual strain not exceeding 0.01\%.

High-resolution TEM (HR-TEM) was carried out further to study the inner structure and composition of the Au-Si MSs (Samples \#1 and \#3). Focused ion beam (FIB) milling procedure was undertaken for isolated MSs to prepare their nm-thick lamellae whose representative HR-TEM image is provided in Figure 2b along with superimposed EDX elemental maps. Both images shed light onto a complex inner structure of the product where Au material fills in the space between irregularly shaped Si nanocrystals. The crystalline structure of Si inclusions was confirmed by HR-TEM images and by single-particle Raman spectroscopy (see Section 2.2). Along with single Au nanoparticles located on the surface of Au-Si MSs, a nm-thick oxide shell can be observed all around the entire Au-Si microsphere (Figure 2b,c). A similar oxide shell was also found in the product obtained at the smallest amount of HAuCl$_4$ added to the laser-processed dispersion (Sample \#1; Figure S3, ESI). To study the inner structure of the Au-Si hybrids more systematically, a series of consecutive FIB cuts of an isolated MS was prepared, while its 3D model was then reconstructed from a series of corresponding SEM images (see Methods and Figure 2d). The performed studies revealed that Si nanocrystals are mainly located near the MS's surface, while its core was mostly formed by Au-based phase.

\begin{figure*}[t!]
\centering
\includegraphics[width=1.5\columnwidth]{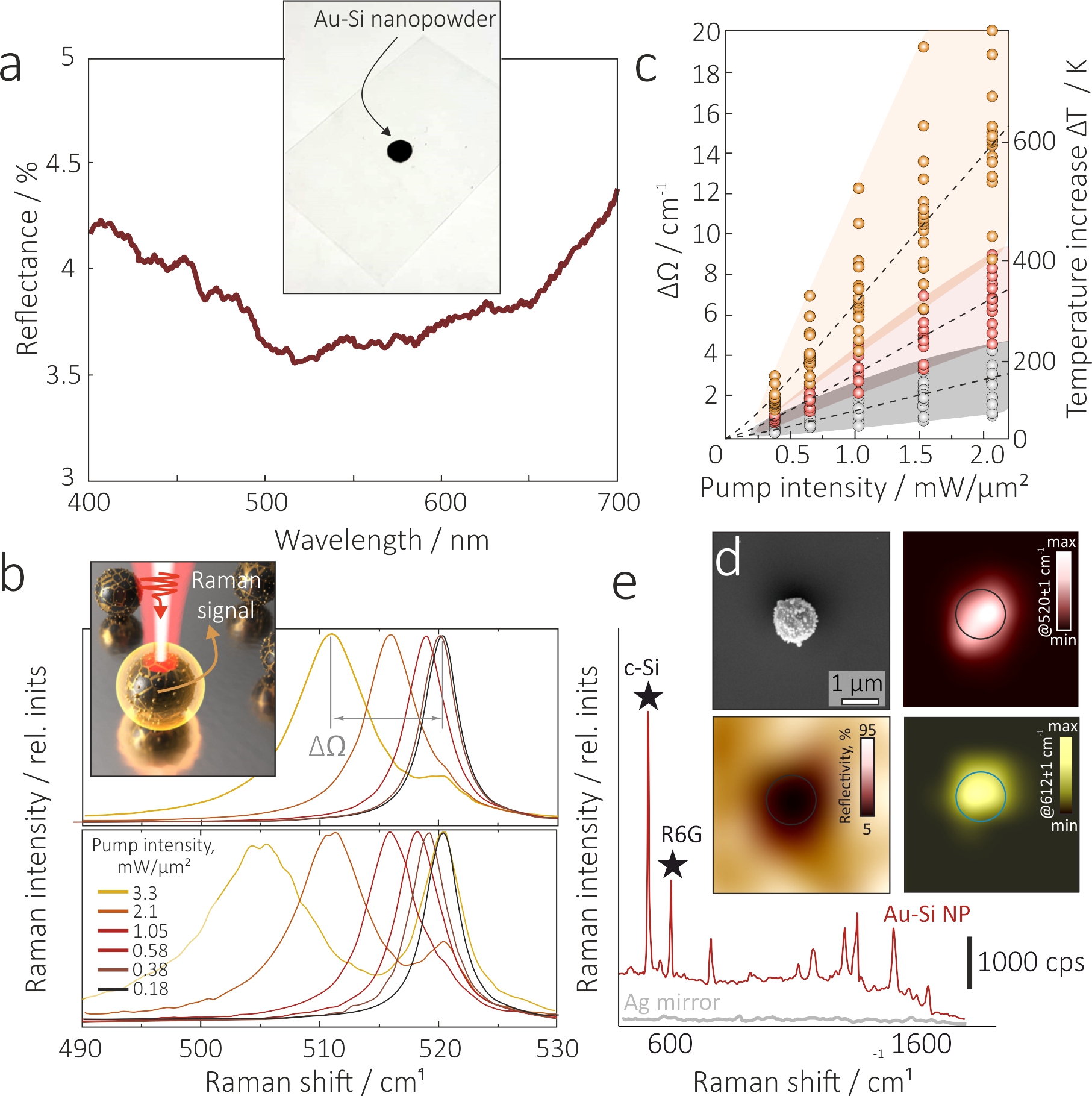}
\caption{\textbf{Light-to-heat conversion and single-particle SERS sensing with Au-Si hybrids.} (a) Reflectance spectrum of Au-Si MS coating. Inset: Optical photograph of coating made of Au-Si MSs (Sample \#2) dried over a glass slide. (b) Representative series of normalized Raman spectra (c-Si band) of separate Au-Si MSs (Samples \#2 and \#3) measured at varied intensity of pump laser radiation $I$. (c) Spectral shift of the c-Si Raman band $\Delta\Omega$ and the average MS temperature increase $\Delta$T versus pump laser intensity $I$ measured for Au-Si MSs from Sample \#1 (gray), Sample \#2 (red), and Sample \#3 (yellow). (d) Correlated SEM, reflection and SERS (at 520.8$\pm$1 and 612$\pm$1 cm$^{-1}$) images of a single Au-Si MS functionalized with R6G molecules. (e) Averaged SERS spectra of the R6G detected from the Au-Si MS and smooth Au mirror sites. Characteristic bands related to crystalline Si and R6G used for mapping are marked by stars. Circle indicates the geometric position of the MS in the optical and Raman images.
}
\label{fig:3}
\end{figure*}

The unveiled 3D morphology of the Au-Si MSs produced at higher amounts of HAuCl$_4$ allowed us to make the following justification regarding the physical and chemical processes underlying the hybrid structure formation. Keeping in mind the transparency of isopropanol at 532 nm, the formation of Au nanoparticles via thermally-induced reduction of HAuCl$_4$ is expected to be preferentially facilitated near the interface of precursor Si particles that efficiently absorbed laser radiation. Formation of Au nanoparticles on the Si surface further boosts the light-to-heat conversion via plasmon-mediated enhancement of the electromagnetic field at 532 nm (see Figure S4, ESI). Then, taking into account the comparable melting temperatures of Au and Si, this may result in remelting of Si surface and capping the formed Au nanoparticles. The molten Au can form a continuous nm-thick coating (Figure S3, ESI) or penetrate between Si nanocrystallites formed under fast recrystallization. The latter effect is associated with weak solubility of Au in Si and was previously predicted by molecular dynamic modeling \cite{larin2020plasmonic}. Subsequent irradiation by laser pulses causes multiple melting/resolidification cycles for highly absorbing Au-Si MSs and formation of Au nanoparticles on their surface.

\subsection{Optical sensing and anti-counterfeiting with Au-Si material}

Rational combination of Au and Si elements within a hybrid structure with a well-developed surface morphology permits to create functional nano/micro-materials with linear and nonlinear optical properties attractive for various applications. First of all, combination of plasmon-active nanomaterial with an indirect bandgap semiconductor provides conditions for efficient light absorption. Reflection spectra of the prepared Au-Si powder  (Sample \#2) drop-cast on a glass slide indicates its average reflectivity $\approx$4 \% in the visible spectral range with the minimal value ($\approx$3.5\%) achieved near the plasmonic absorption band of Au constituents (Figure 3(a,b)). The absorbed laser energy is converted to heat, while the efficiency of this process can be assessed by tracing the spectral position of the main phonon-mediated Raman band of crystalline Si (at 520.8 cm$^{-1}$) \cite{zograf2017resonant,mitsai2019si}. The spectral shift $\Delta\Omega$ of this band can be recalculated to local temperature increase $\Delta$T using the following expression: \cite{balkanski1983anharmonic}

\begin{equation}
\Delta\Omega(T)=\Omega_0+A(1+\frac{2}{e^x-1})+B(1+\frac{3}{e^y-1}+\frac{3}{(e^y-1)^2})
\end{equation}

\noindent where $\Omega_0$=528 cm$^{-1}$, A= -2.96 cm$^{-1}$, A= -0.174 cm$^{-1}$, $x$=$\hbar\Omega_0$/2$\kappa$T and $y$=$\hbar\Omega_0$/3$\kappa$T.

Two representative series of Raman spectra measured at different laser pump intensities from isolated MSs randomly chosen from Samples \#1 and \#3 (top and bottom plots in Figure 3b) demonstrate the remarkably different efficiency of their laser-induced heating. For example, upon excitation at 3.3 mW/$\mu$m$^2$ (maximal laser intensity available in these experiments) the MS containing less Au (Sample \#1) can be heated up to about $\Delta$T$\approx$200 K, while a similar excitation of the Au-rich MS (Sample \#3) resulted in a trice larger temperature increase $\Delta$T$\approx$600 K (Figure 3c). Systematic studies performed for multiple isolated Au-Si MSs produced at varied  HAuCl$_4$ content in the processed dispersion (Samples \#1-\#3) confirm the direct relationship between Au content and light-to-heat conversion efficiency (Figure 3c). Meanwhile, further increase in Au content within the MS is expected to decrease their heating efficiency.
Importantly, even for highly heated hybrids the spectrum of their phonon-mediated Si Raman band was found to preserve a Gaussian-shaped profile, thus  indicating their uniform heating by laser radiation \cite{aouassa2017temperature}.

\begin{figure*}[t!]
\centering
\includegraphics[width=1.8\columnwidth]{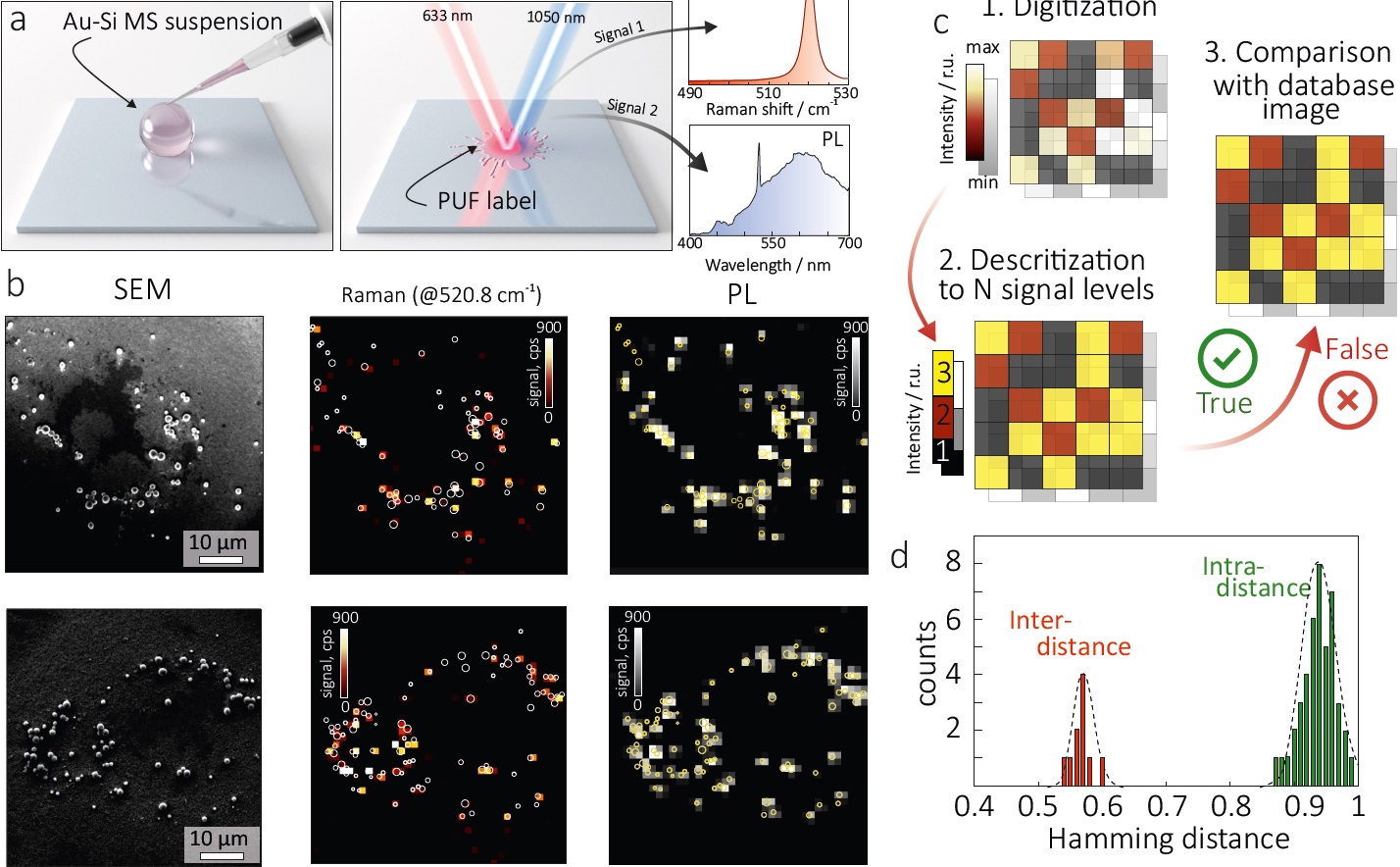}
\caption{\textbf{Optical PUF labeling with Au-Si hybrids.} (a) Schematically illustrated fabrication process of PUF labels and their optical readout by mapping characteristic optical fingerprints: Raman signal and nonlinear broadband photoluminescence associated with crystalline Si inclusions. (b) Correlated SEM, Raman (at 520.8 cm$^{-1}$) and nonlinear PL images of two representative labels composed of random arrangements of the Au-Si MSs. (c) Sketch of label digitization and authentication. (d) Hamming intra- and inter-distances estimated for 10 different PUF labels.
}
\label{fig:3}
\end{figure*}

Inherent Raman signal enhanced by efficient light absorption and plasmon-mediated electromagnetic fields near the Au-Si interfaces make the LAL-prepared MSs promising for nano-sensing based on surface-enhanced Raman scattering (SERS) and photoluminescence (SEPL) effects. The complex chemical composition of the generated structures (Au, Si, and SiO$_2$)  can significantly expand the range of molecules that can functionalize their surface using the toolbox of synthetic organic chemistry \cite{linford1993alkyl}. The temperature-feedback Raman signal can be used to maintain temperature in the analyte-MS system to avoid excessive heating of attached organic molecules, most of which are rather prone to thermal decomposition. Also, the Raman signal of Si permits easily to correlate a certain surface site with the MS and corresponding signal from analyte molecules, thus ensuring more reliable SERS measurements. To illustrate nano-sensing modality of the prepared Au-Si material acting as SERS-active structures, their surface was functionalized with Rhodamine 6G molecules (see Methods). The MS concentration in the dispersion was optimized to yield in deposition of isolated MSs on a smooth Au mirror upon drop-casting. Figure 3d provides correlated SEM, laser confocal (reflectivity at pump laser wavelength of 633 nm) and SERS (intensity at 520.8$\pm$1 and 612$\pm$1 cm$^{-1}$) image of a single 1.1-$\mu$m-sized Au-Si MS confirming its ability to detect adsorbed analyte molecules at pump intensity ($\approx$ 0.3 mW/$\mu$m$^2$) while preserving the sensor temperature below that of the analyte's thermal decomposition \cite{qiu2016thermochemical}. At such pump intensity, the R6G Raman signal measured from a smooth Au reference is indistinguishable over the noise level. Meanwhile the laser pump, which is several orders of magnitude higher by intensity, provides a non-zero signal from the R6G on a smooth Au film allowing for systematic measurements to estimate reliably the SERS enhancement factor (EF) for a single Au-Si MS of $\approx$ (6.1$\pm$2.7)$\times$10$^4$ as a ratio of signals of the main R6G Raman bands normalized to collection time and laser pump intensity (see Figure S5, ESI). Such an EF value can be largely attributed to the plasmon-mediated enhancement of the electromagnetic field amplitude as chemical enhancement is expected to act equally for both Au-Si MS and Au reference. Considering that chemical EF is estimated to be $\approx$10-10$^2$, the total SERS EF factor of the isolated hybrid Au-Si particle is varied between $\approx$10$^6$-10$^7$, which meets the modern requirements for SERS-active substrates \cite{langer2019present}. In particular, the hybrid MSs permit to combine the benefits of plasmonic and nanophotonic approaches for efficient hot-spot engineering with the ability of versatile liquid-phase functionalization and physical analyte concentration in the process of deposition \cite{lee2019designing}.

Mixing nanocrystalline Si with plasmon-active materials is known to provide promising strategy to boost non-linear optical response of such hybrid nanomaterials\cite{knight2011photodetection,cho2013silicon,makarov2018nanoscale,larin2020plasmonic}. In particular, injection of photo-generated hot electrons from bulk Au inclusions and from surface Au nanoparticles into Si nanocrystals upon excitation with femtosecond IR laser pulses was previously shown to provide conditions for observation of nonlinear broadband visible-light PL from the LAL-generated Au-Si nanomaterial \cite{gurbatov2022hybrid}. In the present work, we carried out systematic studies revealing that the highest nonlinear PL yield does not correlate with the amount of Au species added to the laser-treated dispersion (see Methods). This probably indicates the existence of an optimal ratio between the amount of light-emitting Si nanocrystals in the top hemi-sphere of a laser-pumped MS and the light-absorbing Au fraction therein. In particular, the Au-Si MSs produced at moderate HAuCl$_4$ content (Sample \#2) demonstrated the highest integral PL signal (see Figure S6, ESI).

The coexistence of nonlinear PL and characteristic Raman signal within Au-Si MSs makes them quite a unique nanomaterial that can be used to create state-of-the-art optical labels for anti-counterfeiting applications based on physically unclonable function (PUF) strategy \cite{pappu2002physical,mcgrath2019puf}. The PUF approach suggests introduction of some stochastic processes into the label fabrication, thus making them almost unclonable. Optical PUF labels (i.e labels with a read-out based on a certain optical signal such as Raman-scattering, PL and reflection \cite{arppe2017physical}) created using various light-emitting materials have recently gained a lot of research interest owing to simplicity of realization (for example, $via$ simple drop-casting onto the substrate) and authentication. Organic-based light emitters suffer from degradation caused by UV light or elevated temperature, which is why inorganic nanomaterials with unique optical response are beneficial for realization of robust labels. Moreover, the commonly criticized randomness of  LAL-produced nanomaterials (and their related properties), if combined with the randomness of drop-cast deposition from dispersion and agglomeration on the surface, are expected to provide unique conditions for the fabrication of physically unclonable tags that cannot be replicated by chance or with any deterministic approach.

Here, we produced security tags by drop-casting 0.5-$\mu$L drops of as-generated isopropanol dispersion with Au-Si MSs on glass slides or Au mirrors (Figure 4a). Upon drying, the drop-cast material created random arrangements on the surface which are impossible to replicate by chance upon subsequent deposition cycles. Meanwhile, the wafer usually demonstrates very good wettability with respect to isopropanol, which results in spreading and deposition of material over centimeter-sized surface area. Consequently, this increases the material consumption and makes either visual or optical microscopy analysis of the deposition area rather complicated. Typically, a certain randomly chosen area within a relatively large deposition area (in the form of so-called ``coffee rings'') is analyzed, which makes finding and recognition of the tags an extremely hard task for the end user (in absence of additional marking) \cite{gu2020gap}. To avoid the above-mentioned issues, a certain amount (about 5 - 10 vol. \%) of distilled water was added to the as-prepared isopropanol dispersion, which remarkably changed its wetting and drying behavior. We found experimentally that dispersions with different isopropanol/water ratio spread and evaporated at different rates, which allowed us to find conditions when the MSs tended to concentrate within a small droplet that evaporated over a smaller area. Eventually, tags with sizes as small as $\approx$80$\times$80 $\mu$m$^2$ were produced by means of an ordinary syringe, which was achieved through the optimization of the isopropanol/water ratio in drop-cast droplets and concentration of the Au-Si MSs (Figure S7, ESI). Typical examples of Au-Si MS deposits are illustrated by the SEM images in Figure 4b.

The produced tags can be characterized by using certain unique optical fingerprints, for example, their Raman signal (at 520.8$\pm$2 cm$^{-1}$) associated with nanocrystalline Si inclusions in the Au-Si hybrids and/or a related nonlinear broadband PL signal. In Figure 4b, corresponding Raman and nonlinear PL maps (40$\times$40 pixels) reveal close correlation between the real position of MSs (marked by white/yellow circles) and the characteristic distribution of both optical signals. The variation of both signal intensities depends on the local nanostructure of produced Au-Si MSs and their number within each individual pixel as several agglomerated particles can contribute to the overall detected signal. Figure 4c schematically illustrates the authentication/validation procedure of the produced labels. First, coarse-grained discretization of fingerprint signals (either Raman or PL spectra) to a certain number of intensity levels $N$ is performed for each pixel in the Raman/PL map. The end user can repeat optical mapping and leveling procedures validating the PUF label $via$ the pair-wise pixel-by-pixel comparison of signal intensities. Considering the independence of both fingerprint signals, the encoding capacity for the designed tags based on the novel hybrid Au-Si material can be assessed as $N^{2\cdot m}$ (where $m$ is a total number of pixels in the obtained maps). For the illustrated maps (40$\times$40 pixels) the corresponding encoding capacity reaches $\approx$10$^{1500}$ at $N$=3 intensity levels for both characteristic signals, which meets the requirements for modern anti-counterfeit labeling. To justify the ability of the proposed algorithm reliably to distinguish different labels, we characterized 10 random realizations of PUF tags calculating their similarity $\alpha$ (the ratio of zero-intensity pixels to the total pixel number with initial non-zero intensity) or Hamming intra-distance (1 - $\alpha$/100\%). The obtained results are presented in Figure 4d, giving the average inter-Hamming distance of $\approx$0.94$\pm$0.05 for different PUF labels analyzed by mapping both (Raman and PL) signals. Next, each label was scanned one more time and pair-wise comparison of the digitized maps was carried out between the scans of the same label, revealing their strong similarity with the average Hamming intra-distance of $\approx$0.55$\pm$0.04 (Figure 4d). This finding indicates the stability of the developed algorithm and its capability to reliably distinguish between the real (stored in the database) and fake (duplicated) labels.

\section{Conclusion and Outlook}
To conclude, this study highlights the potential of a scalable laser ablation in liquid combined with laser irradiation to create functional micro/nano-materials in which such contrasting constituents as Si and Au can be efficiently mixed within unified structures that exhibit promising optical properties. The combination of plasmon-active Au and nanophotonic Si was found to be useful for the realization of reliable SERS sensors that provide, for isolated hybrid Au-Si structures, an average enhancement factor as high as 10$^7$ and permit to maintain the local temperature of the sensor system below that of analyte's thermal decomposition. The light-to-heat conversion efficiency of the novel hybrid structures was controlled by tuning their Au content, which was confirmed by Raman spectroscopy. Importantly, our preliminary studies also showed the ability to downscale the average diameter of the hybrid Au-Si MSs by using pure Si precursor with smaller diameter (see Figure S8, ESI).

In addition, the commonly criticized randomness of LAL-prepared nanomaterials was utilized in this study to produce physically unclonable anti-counterfeiting labels with encoding capacity of 10$^{1500}$. Larger encoding capacity of the PUF labels can be easily achieved by increasing the number of pixels analyzed in Raman/PL maps or by functionalizing the hybrid Au-Si product with Raman-active molecules capable of providing additional bands for mapping \cite{gu2020gap}. The PUF labels were produced by optimized drop-casting of the as-prepared Au-Si dispersions. The ability to concentrate produced MSs is very advantageous for SERS applications as well as for adequate characterization of the nanomaterial on the stage of synthesis optimization, especially when the droplet contains a small amount of analyzed nanoparticles.

\section{Materials and Methods}
\subsection{LAL-based preparation of hybrid Au-Si structures}
The dispersion of pure polycrystalline Si MSs was first produced by ablating bulk monocrystalline Si wafer placed in isopropanol. We used second-harmonic (532 nm) laser pulses generated with a pulse width of 7-ns and repetition rate of 20-Hz by a Nd:YAG source (Ultra, Quantel). Laser pulses with a constant pulse energy of 0.4 mJ were focused on a Si wafer surface by a lens with a focal distance of 20 mm. Then, a certain amount (ranging from 0.25 to 0.75 mL) of aqueous HAuCl$_4$ (10$^{-3}$M) was added to the as-prepared Si dispersion, after which the mixture (6 mL in volume) was further irradiated for another 1 h keeping the same laser parameters.

\subsection{Characterization}
Initial assessment of the morphology and composition of the prepared hybrid MSs was carried out using a SEM tool (Ultra 55+, Carl Zeiss) equipped with an EDX detector (X-max, Oxford Instruments). EDX measurements were performed at accelerating voltage of 5 kV.

Absorbance spectra of the dispersions containing LAL-generated Si nanoparticles and the final hybrid Au-Si structures (Samples 1-3) were recorded using a UV–Vis spectrophotometer (UV-2600, Shimadzu) equipped with an integrating sphere (ISR-2600 Plus). The measurements were carried out at room temperature using a quartz cuvette and absorbance spectrum of pure isopropanol as reference. Reflection spectra of the dry powder of Au-Si MSs (Sample \#2) were obtained using a spectrophotometer (Cary Varian 5000) equipped with an integrating sphere.

 XRD patterns of hybrid Au-Si MSs on Si wafer were collected on a diffractometer with CuK$_{\alpha}$ radiation operated in the 2$\Theta$/$\omega$ mode (SmartLab, RIGAKU). The incident beam size did not exceed 0.1 mm, and the XRD peaks were identified in accordance with the ICDD database.

Imaging of the MSs combined with their focused-ion beam (FIB) milling modality was performed with a SEM instrument (Helios 450s Nanolab, Thermofisher). FIB milling permitted to prepare precise cross-sectional cuts of isolated MSs which were then examined by SEM. By repeating milling procedure, we realized the FIB tomography to create a 3D reconstruction of the internal and external features of the prepared MSs. Series of slices were obtained automatically via the FEI AutoSlice \& View G3 software (Thermofisher). The MS surface was protected by a locally deposited Pt layer. Ion beam was operated at the energy of 30 keV and a beam current of 24 pA. The total number of slices was about 110 per one MS, with an approximate slice thickness of 7 nm. SEM image acquisition was set up for recording images with 1536 x 1024 pixels (with pixel size being 1.3 nm x 1.3 nm). As a result, the 3D volume of 2 $\mu$m x 1.3 $\mu$m x 0.7 $\mu$m was collected. Further, the resulting volume was processed with the Avizo 8.1 software (Thermofisher). FIB milling was also used to prepare thin samples for TEM studies. HR-TEM images were acquired at the acceleration voltage of 80 kV (Titan 80-300, Thermofisher).

Modeling of the normalized electromagnetic field amplitude ($E$/$E_0$) near a pure and Au-decorated Si MSs was carried out using the Comsol Multiphysics package. We considered excitation of an isolated particle in isopropanol (refractive index of 1.385) by a linearly polarized plane wave with a wavelength of 532 nm. Temperature profiles under single-pulse excitation of the MS were then calculated by numerically solving the heat equation using the same software package.

\subsection{Raman thermometry, sensing and PUF labeling}
Raman micro-spectroscopy was carried out using a multifunctional optical system (Spectra II, NT-MDT) equipped with a CW He-Ne laser (central wavelength of 633 nm), a femtosecond-pulse IR laser (TOPOL-1050-C, Avesta Project; pulse repetition rate of 80 MHz, central wavelength of 1050 nm) and grating-type spectrometer with a monochromator (M522, Solar Laser Instruments) and electrically cooled CCD camera (i-Dus, Oxford Instruments). Pump laser radiation was focused on the sample surface with a dry microscope objective (100$\times$, M Plan Apo, Mitutoyo) having a numerical aperture of 0.7. Light-to-heat conversion measurements were carried out with isolated Au-Si MSs drop-cast on a monocrystalline Si wafer. Raman maps (40$\times$40 pixels) of the PUF labels were recorded from 80$\times$80 $\mu$m$^2$ surface areas at pump intensity of 0.5 mW/$\mu$m$^2$ and signal accumulation time of 0.1 per pixel. Correlated nonlinear PL maps (40$\times$40 pixels) were acquired by scanning the same surface area with fs-laser radiation at a pulse energy of $\approx$ 0.8 nJ and collecting the integral PL intensity within the visible spectral range (400 - 700 nm).

R6G molecules were used  as a probe for the assessment of SERS performance of the Au-Si hybrids owing to its strong affinity to noble metals and high Raman cross-section. The Au-Si MSs (Sample \#2) were functionalized with R6G molecules by mixing a freshly prepared MS dispersion in isopropanol with aqueous R6G (10$^{-6}$ M). The obtained mixture was kept for 1 h and then drop-cast on a bulk Au mirror reference. A similar analyte functionalization procedure was carried out for Au reference to evaluate the SERS EF. The PUF labels were created by drop-casting droplets (0.5 $\mu$L) of  as-produced Au-Si dispersion on Ag mirror/glass slides using a conventional syringe.

\section*{Supporting Information}
There are no conflicts to declare.

\section*{Acknowledgements}
This work was supported by the Russian Science Foundation (Grant no. 21-79-00302). S.G. expresses his gratitude to the Ministry of Science and Higher Education of the Russian Federation (Grant Nos. MK-4321.2021.1.2).

\section*{Conflicts of interest}
The authors declare no conflict of interest.

\section*{Data Availability Statement}
The data that support the findings of this study are available from the
corresponding author upon reasonable request.

\section*{Keywords}
Laser ablation in liquids; Hybrid nanomaterials; Au-Si microspheres; SERS; Optical labels

%

\end{document}